# Phonon Scattering Mechanism in Thermoelectric Materials Revised via Resonant X-ray Dynamical Diffraction

**Adriana Valério, Rafaela F. S. Penacchio, Maurício B. Estradiote, Marli R. Cantarino, Fernando A. Garcia, and Sérgio L. Morelhão,** Institute of Physics, University of São Paulo, São Paulo, SP, Brazil

**Niamh Rafter and Stefan W. Kycia,** Department of Physics, University of Guelph, Guelph, Ontario N1G 1W2, Canada

**Guilherme A. Calligaris,** Brazilian Synchrotron Light Laboratory - LNLS/CNPEM, Campinas, SP, Brazil

**Cláudio M. R. Remédios,** Instituto de Ciências Exatas e Naturais, Universidade Federal do Pará, Belém, PA, Brazil

Address all correspondence to Sérgio L. Morelhão at morelhao@if.usp.br

## Abstract

Engineering of thermoelectric materials requires an understanding of thermal conduction by lattice and electronic degrees of freedom. Filled skutterudites denote a large family of materials suitable for thermoelectric applications where reduced lattice thermal conduction attributed to localized low-frequency vibrations (rattling) of filler cations inside large cages of the structure. In this work, a multi-wavelength method of exploiting X-ray dynamical diffraction in single crystals of $CeFe_4P_{12}$ is presented and applied to resolve the atomic amplitudes of vibrations. The results suggest that the vibrational dynamics of the whole filler-cage system is the actual active mechanism behind the optimization of thermoelectric properties.

## Introduction

Understanding a mechanism for impeding thermal conductivity, while keeping a high electric conductivity, is key to design better thermoelectric materials.[1-10] Filled skutterudites are a family of materials displaying good potential for thermoelectric applications.[11-18] This is usually attributed to an enhancement of phonon scattering, which lowers the thermal conductivity, by localized vibrational modes denoted rattling modes.[19-21]

The filled skutterudites have the general formula $RT_4X_{12}$, where the $T_4X_{12}$ atoms form a large icosahedral cage inside which the filler $R$ atom resides. In the simplest description, the rattling vibrations are attributed exclusively to the filler. This partitioning of the skutterudite vibrational dynamics into the filler and cage subsystems is believed to provide a good description of a large set of skutterudites.[11,20,22]

The synthesis of materials also have significant impact over the thermoelectric properties as grain boundaries are known to play a role in the phonon scattering and thermal conductivity.[13-18,23] Although it is feasible to solve the microstructures of materials,[24,25] the intrinsic lattice thermal conductivity is better investigated in single crystals[26]. Filled skutterudites can be synthesized in the form of high quality single crystals,[12,27,28] inviting the perspective of revisiting the phonon scattering mechanism through X-ray phase measurements via dynamic diffraction effects.[29-35]

Standard methods in X-ray crystallography rely on the intensity data of many individual reflections and the structure determination is based on the best fit of the whole available data set. The reliability of the results depends upon the accuracy by which structural factor modules were extracted from the diffracted intensities.[36] The phase measurement method is a completely different approach which explores the effects of interference between diffracted waves when more than one Bragg reflection is excited. The information extracted from the phases of the structure factors is then used to validate or select feasible model structures for the crystal.[37-39] The minimum crystal size for this type of experiment is, in general, ten times larger than the usual grain size found in powder samples.[40]

Structure factor phases are susceptible to the differences between the vibration amplitudes of the atoms—root mean square (RMS) atomic displacements. In other words, phase values are invariant, as a function of temperature, only when all occupied sites of the unit cell have equal values of Debye-Waller factors. In this work, structure factor calculations in model structures revealed suitable Bragg reflections and X-ray energies to resolve the difference in atomic vibrations. It also revealed a giant abrupt resonant phase shift for the whole family of filled skutterudites $RFe_4P_{12}$ (R=Ce, La, Nd, Pr, Sm).[41,42] To exploit this resonant phase shift, multi-wavelength data collection and analysis procedures were developed and applied to a single crystal of $CeFe_4P_{12}$.[43,44] Preliminary results point towards a scenario where rattling of the filler atoms Ce alone is not enough to explain the $CeFe_4P_{12}$ vibrational dynamics, suggesting that a complex interaction between the filler-cage subsystems is a common thread to many skutterudites.[21]

## Experimental details

The starting materials were Fe and Ce powders (99.99%, American Elements), red P (>99.99%, Sigma-Aldrich Chemical Co.), and Sn powder (99.999%, American Elements). Single crystals were grown by the tin-flux method.[43,44]

Checking of crystalline perfection for dynamical diffraction and lattice parameter determination at room temperature, 297 K, were carried out with characteristic radiation in a Huber four-circle diffractometer sourced by a fine focus copper rotating anode configured with a double collimating multilayer optics followed by a double bounce Ge 220 channel-cut monochromator. Bandwidth is 2 eV for CuK$\alpha_1$ ($\lambda$=1.540562 Å). The reading of the diffracted intensity was performed by a sodium-iodide scintillation detector. Adjustment arcs, Fig. 1(a), of the goniometer head were used to set the diffraction vector of reflection 002 collinear with the $\Phi$ rotation axis of the diffractometer with an accuracy better than $0.01°$. Positive rotation sense of the $\Phi$ axis is clockwise, and its zero defined when the [110] direction is in the diffraction plane pointing upstream. Azimuthal scans of good angular resolution are possible in this diffractometer because of the narrow axial (vertical) divergence of about 0.015°, which is only three times the horizontal divergence of 0.005°.[45,46] The beam size at the sample position was trimmed down to 1×1 mm².

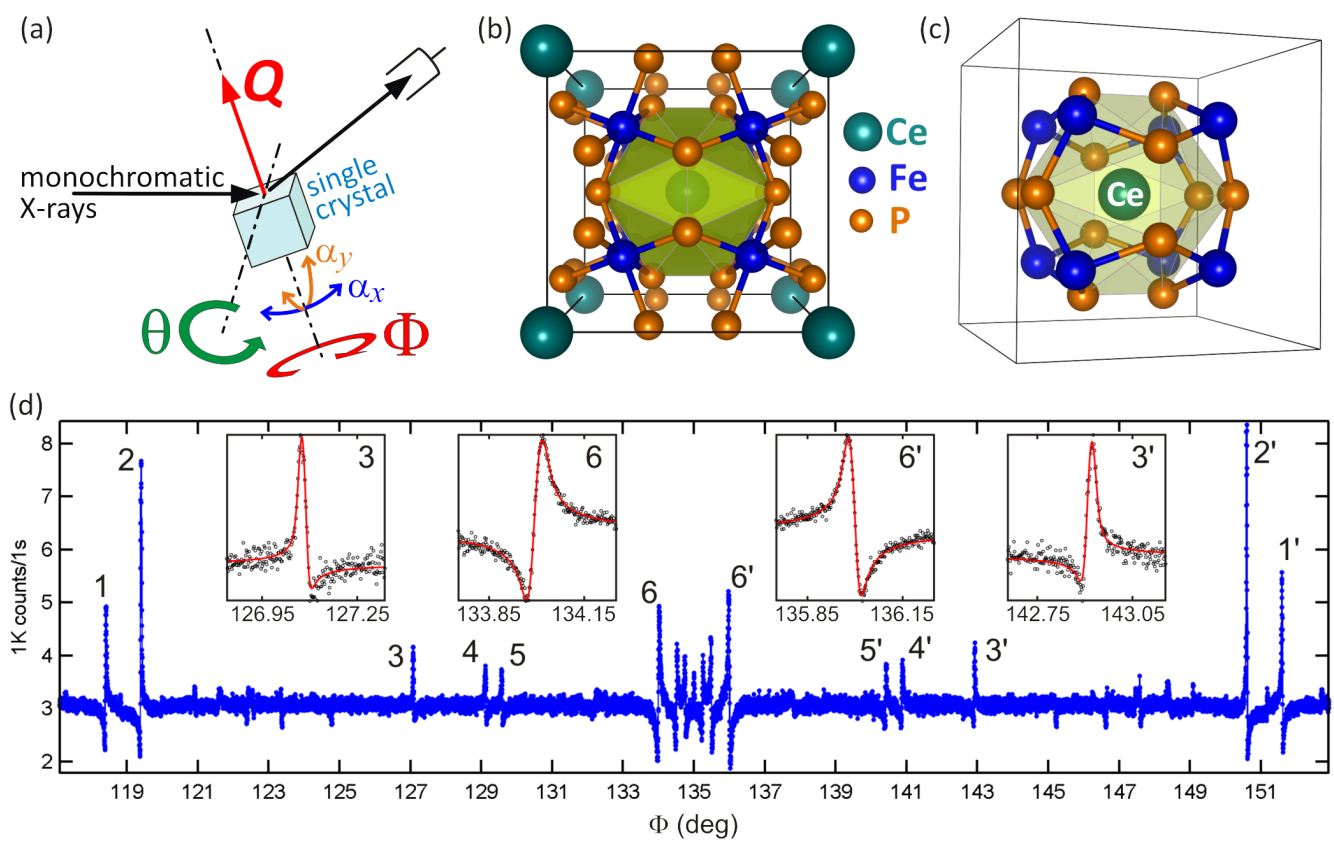

**Figure 1.** (a) Basic goniometry for azimuthal $\Phi$ scan of a Bragg reflection with diffraction vector **Q**. Adjustment arcs $\alpha_x$ and $\alpha_y$ of the goniometer head used to set vector **Q** collinear with rotation axis $\Phi$, and rotation axis $\theta$ used to keep vector **Q** undergoing diffraction. (b) Cubic unit cell of CeFe$_4$P$_{12}$ containing eight Fe$^{2+}$ filled P octahedra, P-Fe distance 0.2248 nm. (c) Ce ion inside the icosahedral cage, P-Ce distance 0.2996 nm. Lattice parameter a = 0.77918±0.00002 nm at room temperature. (d) $\Phi$ scan of the 002 reflection, CuK$\alpha_1$ radiation (8048 eV, in-house diffractometer). Numbers with quotes stand for symmetry related peaks. Peak indexing is available as Supplementary Material. Asymmetric line profiles of a few peaks are detailed in the insets.

Resonant X-ray diffraction were carried out at the Brazilian Synchrotron Light Laboratory (LNLS), bending magnetic beamline XRD2. The beam was vertically focused with a bent Rh-coated mirror, which also filtered higher-order harmonics. Energy was tuned using a double-bounce Si (111) monochromator with sagittal second crystal, placed after the Rh mirror. X-ray optics were in parallel-beam mode (mirror and sagittal crystal focused at infinity): spectral resolution of about 5 eV ($\Delta E/E = 8 \times 10^4$), divergences of 0.1 mrad (0.006°), and beam sizes of 0.5 mm at the sample position. The sample was mounted onto the Eulerian cradle of a Huber 4+2 circle diffractometer in the same orientation used before for the in-house measurements, that is the azimuthal scan of reflection 002 carried out by using the basic goniometry represented in Fig. 1(a). X-ray diffraction data were collected at σ-polarization (vertical scattering plane) by a Pilatus 100 K area detector: diffraction spot intensity as the counting rate on a few pixel area.

## Results and discussion

Asymmetric intensity profiles of *n*-beam diffraction peaks are the most undeniable evidence of dynamical diffraction, exactly as seen in the azimuthal scan presented in Fig. 1. The base line intensity is provided by the reference reflection always in diffraction condition during the Φ rotation of the crystal around the diffraction vector of the reference reflection, Fig. 1(a). When another reflection is brought into diffraction condition by the crystal rotation, a secondary wave is produced that can interfere with the reference wave. As these waves came from distinct sequences of reflections, they have different phases. In perfect crystals where phase coherence is not compromised by lattice defects, interference effects between these waves are observable as described by the dynamical theory of X-ray diffraction.[30,47]. The secondary wave undergoes a 180° phase shift across the diffraction peak, then constructive (destructive) interference on one side becomes destructive (constructive) on the other side, producing the observable asymmetries of the n-beam diffraction peaks. The well defined asymmetries seen in Fig. 1(d) are characteristic of highly perfect crystals as semiconductor crystals. Observing asymmetric peaks in azimuthal scans mean that there is accessible information about the phases of the structural factors. Besides being the simplest way to select single crystal samples of enough perfection for phase measurements, azimuthal scanning is well known as the most accurate method to determine lattice parameters.[48-51] By simply measuring the relative peak distances, as between peaks 6 and 6' in Fig. 1(d), accuracy of the order of $10^{-6}$ can be achieved in monitoring relative variations of lattice parameters caused by changes in the sample environment

such as temperature, pressure, strain, or applied electromagnetic fields.[52]

Phonon scattering in skutterudites has been attributed to localized vibrations of the filler,[20] Ce in this case, inside a huge icosahedral cage. Diffracted X-ray waves have amplitude and phase given by the structure factors

$$F(Q)=\sum_n f_n \exp(-M_n)\exp(iQ \cdot r_n)=|F(Q)|\exp[i\delta(Q)]$$

where $r_n$ and $f_n$ stand for positions and scattering amplitudes[53] of the atoms in the unit cell. The Debye-Waller (DW) factor $M_n=\frac{1}{2}Q^2\langle|(r_n-\langle r_n\rangle)\cdot\hat{Q}|^2\rangle=\frac{1}{2}Q^2 u_n^2$ is summarized in terms of the RMS atomic displacements $u_n$ along the direction $\hat{Q}$ of the diffraction vector $Q=Q\hat{Q}$. Standard methods of X-ray crystallography rely on the modulus, $|F(Q)|$, of the structure factors, or in other words, on the amplitude of the diffracted waves. But, in perfect single crystals undergoing dynamical diffraction, information on structure factor phases $\delta(Q)$ are also accessible. Then, a general important question is how this information can help us to better understand materials properties. Particularly in this study on phonon scattering mechanism, what is the new information we can access about atomic vibrations, or how the DW factor does affect structure factor phases. By looking at the standard formula of the structure factor, we can easily see that when all elements have nearly the same DW factor, reflection phases —more precisely the phase differences between simultaneously diffracted

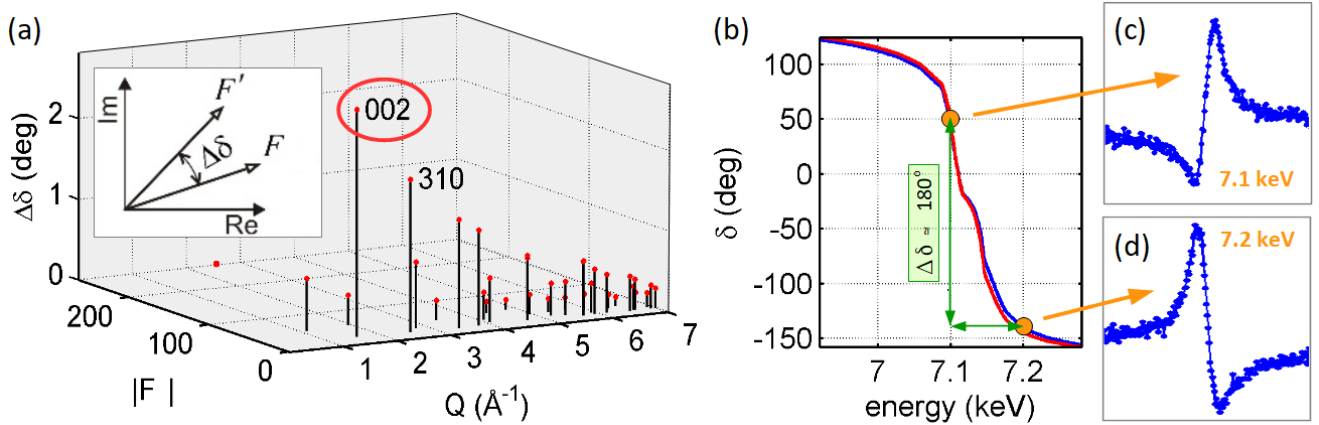

**Figure 2.** (a) Comparison of reflection phases regarding the valence of Ce, changing from 3+ to 4+. The 002 and 310 reflections are the most susceptible ones. (b) Phase of reflection 002 as a function of the X-ray energy near the absorption edge of Fe at 7.112 keV. (c,d) Expected inversion of asymmetry in a n-beam diffraction peak when the phase of the reference wave shifts by 180°.

waves— can be quite invariant with respect to the amplitudes of thermal vibrations. In other words, $|F(Q)|$ is affected by the absolute values of $M_n$, while $\delta(Q)$ is affected only by the relative values between $M_n$ and $M_m$ where $n$ and $m$ stand for the atomic sites of different chemical species.

Reflections with phase susceptible to changes in ionic charges, resonance amplitudes, atomic positions, and occupation and DW factors have been identified by means of model structures.[37-40] For the CeFe$_4$P$_{12}$ structure, the atomic planes of Ce and Fe are interleaved along [001] type directions, Fig. 1(c). As consequence, the 002 reflection phase is the most susceptible one to variations in the scattering amplitudes of these two ions. To illustrate this fact, all reflection phases are compared in Fig. 2(a) for the valence of Ce, between Ce$^{3+}$ and Ce$^{4+}$.[43,44] The 002 reflection displays the largest phase shift $\Delta\delta$, which is small of about $3°$ in this case. However, it is gigantic regarding the resonance amplitudes of Fe. To be more precise, near the Fe absorption edge the phase shift can be as large as $180°$ for just 100 eV of variation in the X-ray energy, Fig. 2(b). Large phase shifts as a function of the X-ray energy allow new strategies to exploit phase measurements. As the phase of the reference wave shifts by $180°$, each *n*-beam diffraction peak undergoes inversion of asymmetry such as seen between the peaks in Figs. 2(c) and 2(d). It means that in the CeFe$_4$P$_{12}$ crystal, by taking the 002 as the reference reflection, the phase of all secondary waves can be determined by monitoring the energy in which each diffraction peak has its asymmetric aspect inverted.

The gigantic phase shift of reflection 002 occurs because the X-ray scattering from the atomic planes of Ce and P nearly cancel the scattering from the atomic planes of Fe (see Argand diagrams in Supplementary Material). Therefore, this effect is also expected in other skutterudites with cage framework Fe$_4$P$_{12}$ and filler ions of the lanthanide family with similar scattering amplitudes of Ce such as La, Pr, Nd, and Sm. A few other reflections such as 222, 280, 820, and 266 also display large phase shifts and, in principle, can also be exploited in this type of experiment, that is $\Delta\delta_{280} \approx \Delta\delta_{002}$, $\Delta\delta_{222} \approx \Delta\delta_{820} \approx 0.5\Delta\delta_{002}$, and $\Delta\delta_{266} \approx 0.36\Delta\delta_{002}$. The 002 was chosen mainly because its phase shift is the largest. Other benefits of this choice are the 4-fold symmetry of the cubic unit cell [001] direction that provides reference positions easy to spot in the $\Phi$-scans, e.g. the mirroring symmetry around $\Phi=135°$ in Fig. 1(d), and the fact that all reflections of this family undergo exactly the same phase shift. For instance, in the case of the 280 reflection an extra procedure would be needed to distinguish it from reflection 820.

Resonant phase shifts have been observed with synchrotron radiation in perfect semiconductor crystals.[33,40] In the CeFe$_4$P$_{12}$ skutterudite, the first experimental confirmation of a huge resonant phase shift of reflection 002 is given in Fig. 3 by means of $\Phi$-scans carried out at room temperature and slightly different X-ray energies: one at 7105.8 eV [Fig. 3(a)] and the other at 7161.2 eV [Fig. 3(b)]. A direct comparison of peak asymmetries between these two scans can be difficult at a first glance as the peak positions are undergoing variations as large as $2°$ for this small difference of 55.4 eV in energy. Then, numbers are used to correlate a few peaks on both scans where numbers with quotes stand for equivalent peaks regarding the crystal space group. Except for the peak at $\Phi=45°$ that is symmetric regardless the X-ray energy, as explained in Supplementary Material, all others have their asymmetric aspect inverted as expected for a phase shift $\Delta\delta$ of the reference wave close to $180°$. It is always useful to emphasize the amazing accuracy of azimuthal scans in determining lattice parameters and/or the used X-ray energy. For instance, the Bragg angle of reflection $16\bar{3}$ change by $\Delta\theta \simeq 0.5°$ with respect to the used energies, while its corresponding peaks in the azimuthal scans change by $\Delta \simeq 2°$, as indicated (arrow A). Moreover, as the diffraction geometry is kept constant during the $\Phi$ rotation, variation in relative peak distance can easily be measured free of instrumental errors.[49,50] At room temperature with Cu$K\alpha_1$ radiation, the obtained lattice parameter was $a=7.7918(\pm0.0002)\text{Å}$. This value has been used to know exactly the X-ray energy during the synchrotron experiments at room temperature in Fig. 3(a) and 3(b). Peak positions changing as a function of temperature can be seen in the $\Phi$-scans performed at the fixed X-ray energy of 7161.2 eV and different temperatures, Fig. 3(b) and 3(c). Lattice parameter $a=7.7862\text{Å}$ is obtained from the $\Phi$-scan in Fig. 3(c), implying in a value of thermal expansion coefficient around $3.8\times10^{-5}\text{Å}/\text{K}$. More reliable values of the thermal expansion coefficient for this material can be achieved in experiments with accurate data of temperature at the sample position, which was not the case here.

    In previous applications of phase measurements, only one X-ray energy was used in each case since simple evaluation of just a few peak asymmetries allowed the validation of the proposed structural models,[37-40] as in the case of detecting hydrogen bonds between amino acid molecules.[39] By properly choosing the reference reflection with graphical help of the two-dimensional (2D) representation of Bragg cones (Supplementary Material),[54] one azimuthal scan at no particular X-ray energy had enough information to evidence the electron charges at the hydrogen bonds. Detecting hydrogen bonds in a single crystal of alanine was a proof of principle, opening new opportunities in terms of fast methods to measure electron charges at hydrogen bonds in biological molecules. Radiation damage to H-bonds is an example of study that can be carried out by exploiting this method. However, lasting of dynamic diffraction regime and lacking of standardized data collection

and analysis procedures stand as the main challenges for phase measurements in crystals of complex biomolecules.

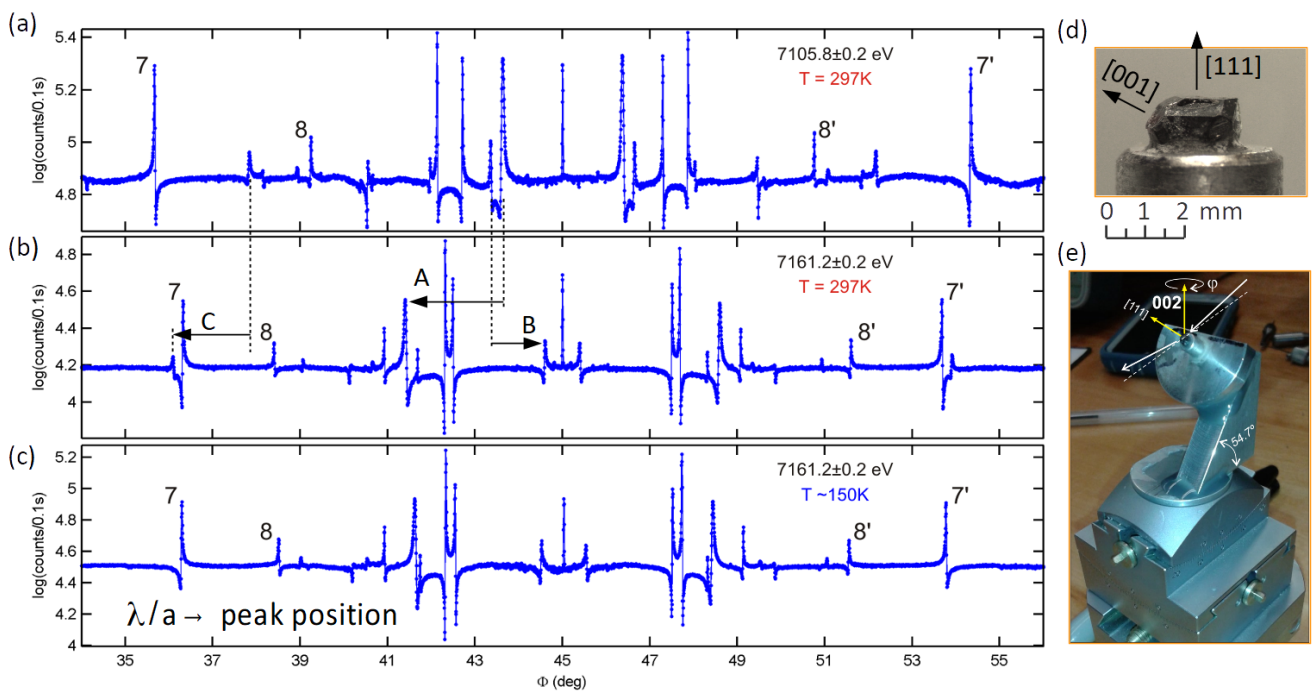

**Figure 3.** (a-c) Azimuthal Φ scans of reflection 002 with synchrotron radiation. X-rays of (a) 7105.8 eV and (b) 7161.2 eV, both at room temperature. Numbers (7, 8, 8', and 7') and arrows (A, B, and C) are used to point out peak position shifting between these two scans. (c) Effect of reducing the temperature from 297 K to near 150 K in the Φ scan with X-rays of 7161.2 eV. (d) Sample of $CeFe_4P_{12}$ used in this work. (e) Assembly of the sample to facilitate setting of direction [001] collinear with the Φ rotation axis.

The multi-wavelength data analysis procedure proposed here consists in creating, for each X-ray energy, N-dimensional arrays containing information on the matching between theoretical and experimental peak asymmetries as a function of the adjustable parameters of the model structures. Each pair of parameters defines a 2D array within a higher dimensional array when taking into account all N adjustable parameters. The number $m$ of measurable asymmetric diffraction peaks from a Φ-scan is represented as sub-arrays of $i \times j \leq m$ binary values such as 1 when there is a match between theoretical and experimental peak asymmetries and 0 otherwise. For instance, nine peaks ($m=9$) can be represented by 3-by-3 sub-arrays, as depicted in Fig. 4(a) where only one mismatch of asymmetry is reported. The sub-arrays are then plotted as a function of two parameters, composing as many as 2D arrays are necessary to go over all possible model structures. The two model parameters investigated here to demonstrate the usage of this analytical procedure in the study of thermoelectric materials are the atomic RMS displacement parameters $u_{Ce}$ and $u_P$ of the $Ce^{3+}$ and $P^{1-}$ ions, respectively. As reflection phases are susceptible to the relative difference between displacement parameters, accounting for variation of $u_{Fe}$ in the models is redundant.

In the complete $\Phi$-scan performed at 8 keV, 68 peaks in the range $\Phi \in [-45°, +45°]$ show reliable asymmetries in terms of the non-overlapping of peaks with conflicting asymmetries.[39] Sub-arrays of 9-by-9 pixels are used in this case where only 68, out of 81, are in fact used to compare peak asymmetries by 1 or 0 values, while the remaining 13 unused positions also display value 1. The asymmetries of these 68 peaks are compared with 121 model structures produced by varying $u_{Ce}$ and $u_P$ from 10 pm to 30 pm in steps of 2 pm. For visual comparison in Fig. 4(b), the 2D array of asymmetry comparison is represented in colors of dark gray (pixel value 1) and orange (pixel value 0). This asymmetry matching diagram (AMD) displays an interesting result, it shows that only models where $u_{Ce}+u_P<36$ pm are compatible with the asymmetries of all 68 peaks. Therefore, measurements at other energies are necessary to resolve the displacement values of these ions.

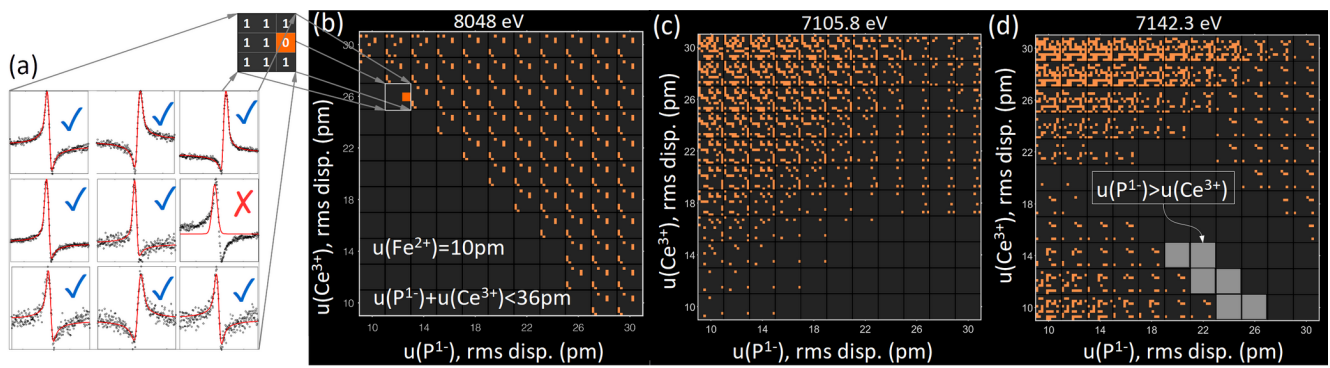

**Figure 4.** (a) Scheme for comparing theoretical and experimental peak asymmetries in terms of arrays of binary 1 and 0 values. (b-d) Asymmetry matching diagrams (AMDs) as a function of RMS displacements of $Ce^{3+}$ and $P^{1-}$ ions. Experimental asymmetries from $\Phi$-scans of reflection 002 with different energies (Supplementary Material): (b) $CuK\alpha_1$ radiation, and synchrotron X-rays of (c) 7105.8 eV and (d) 7142.3 eV. Light-gray squares indicate parameter values providing theoretical peak asymmetries that agree with the entire multi-wavelength data set.

Comparisons of peak asymmetries from synchrotron $\Phi$-scans are shown in Fig. 4(c) and 4(d), AMDs for X-rays of 7105.8 eV and 7142.3 eV, respectively. As a consequence of higher X-ray flux of the synchrotron source, the number of reliable asymmetries increases by at least 25%. Asymmetries underneath the statistical noise when using the in-house setup, Fig. 1(d), are better resolved by the increase of about 10 times in the photon counting rate of the $\Phi$ scans with synchrotron radiation, Fig. 3(a-c). Moreover, there is still the reduction in the relative strength of the 002 reflection near the Fe edge that can also have contributed to the 40% increase in the number of measurable asymmetries with X-rays of 7105.8 eV (Supplementary Material).

When taking into account the previous result in Fig. 4(b), the AMD in Fig. 4(c) states that $u_{Ce} < 16$ pm, while the AMD in Fig. 4(d) sets the lower limit for $u_P$ that can be written as $u_{Ce} + u_P = 35 \pm 1$ pm. In other words, by superposing the three AMDs in Fig. 4(b-d), the only models capable of explain all peak asymmetries for the three different wavelengths are those models highlighted in Fig. 4(d) by light-gray squares. Roughly, the overall result is that $u_{Ce} \simeq 12 \pm 2$ pm and $u_P \simeq 23 \pm 2$ pm when using model structures with fixed $u_{Fe} = 10$ pm or, in more general terms, $u_{Ce} \gtrsim u_{Fe}$ and $u_P \simeq 2 u_{Ce}$ as the peak asymmetries are more susceptible to relative values of atomic displacement parameters.

Lighter P atoms having larger displacement parameters indicates that to keep the 4.5 times heavier Ce inside the icosahedral cage, the neighboring P octahedrons are undergoing random distortion of the intra octahedron Fe-P-Fe bonding angles while keeping the Fe-P distance nearly unchanged. The linear momentum produced by small displacements of Ce is enough to prevent lattice vibration modes involving collective rotation of the octahedrons. Independent rattling of the Ce atoms is therefore in the momentum space. In the real space, the Ce-P coupling induces random vibrations of the P atoms, giving rise to a correlated rattler-cage vibrational dynamics.

## Conclusions

In this work, a huge resonant reflection phase shift was predicted to occur in a family of skutterudites and experimentally observed in a perfect single crystal of CeFe$_4$P$_{12}$. This huge phase shift opened new opportunities for exploiting dynamical diffraction in skutterudite type of thermoelectric materials. In this sense, a general multi-wavelength data collection and procedure was proposed and applied to resolve a simple case based on model structures summarized by only two adjustable parameters. The results indicate the atomic mass ratio between the filler ion and the ions forming the cage as the driven mechanism for phonon scattering.

## Acknowledgments

The authors acknowledge the financial support from Brazilian agencies CAPES (Grant Nos. 88881.119076/2016-01 and 2018-5), FAPESP (Grant Nos. 2018/00303-7, 2019/11564-9, and

2019/15574-9), and CNPq (Grant Nos. 309867/2017-7 and 452287/2019-7), as well as from the NSERC of Canada.

**Supplementary Material:** The supplementary material for this article can be found at https://doi.org/10.1557/mrc.2020.37.